\documentclass[sigconf]{acmart}
\AtBeginDocument{%
  \providecommand\BibTeX{{%
    \normalfont B\kern-0.5em{\scshape i\kern-0.25em b}\kern-0.8em\TeX}}}

\usepackage{cleveref}
\usepackage{gensymb}
\usepackage{listings}
\usepackage{color, colortbl}
\usepackage{multirow}

\setcopyright{acmcopyright}
\copyrightyear{2021}
\acmYear{2021}
\setcopyright{rightsretained}
\acmConference[UbiComp-ISWC '21 Adjunct]{Adjunct Proceedings of the 2021 ACM International Joint Conference on Pervasive and Ubiquitous Computing and Proceedings of the 2021 ACM International Symposium on Wearable Computers}{September 21--26, 2021}{Virtual, USA}
\acmBooktitle{Adjunct Proceedings of the 2021 ACM International Joint Conference on Pervasive and Ubiquitous Computing and Proceedings of the 2021 ACM International Symposium on Wearable Computers (UbiComp-ISWC '21 Adjunct), September 21--26, 2021, Virtual, USA}
\acmDOI{10.1145/3460418.3479326}
\acmISBN{978-1-4503-8461-2/21/09}

\acmSubmissionID{earcomp-5260}

\begin{document}

\title{PilotEar: Enabling In-ear Inertial Navigation}

\author{Ashwin Ahuja}
\email{aa2001@cl.cam.ac.uk}
\orcid{0000-0003-4574-6669}
\affiliation{%
  \institution{University of Cambridge}
  \country{United Kingdom}
}

\author{Andrea Ferlini}
\email{af679@cl.cam.ac.uk}
\orcid{0000-0003-0768-4735}
\affiliation{%
  \institution{University of Cambridge}
  \country{United Kingdom}
}

\author{Cecilia Mascolo}
\email{cm542@cl.cam.ac.uk}
\orcid{0000-0001-9614-4380}
\affiliation{%
  \institution{University of Cambridge}
  \country{United Kingdom}
}

%\renewcommand{\shortauthors}{Ahuja, Ferlini and Mascolo}

\begin{abstract}
Navigation systems are used daily. 
While different types of navigation systems exist, inertial navigation systems (INS) have favorable properties for some wearables which, for battery and form factors may not be able to use GPS.
Earables (\textit{aka} ear-worn wearables) are living a momentum both as leisure devices, and sensing and computing platforms.
The inherent high signal to noise ratio (SNR) of ear-collected inertial data, due to the vibration dumping of the musculoskeletal system; combined with the fact that people typically wear a pair of earables (one per ear) could offer significant accuracy when tracking head movements, leading to potential improvements for inertial navigation.
Hence, in this work, we investigate and propose PilotEar, the first end-to-end earable-based inertial navigation system, achieving an average tracking drift of $0.15 \frac{m}{s}$ for one earable and $0.11 \frac{m}{s}$ for two earables.
\end{abstract}





\maketitle

\section{Introduction}\label{sec:introduction}
Navigation systems are ubiquitous, they are in our phones, cars, sometimes even in our smartwatches.
People rely on navigation systems daily: commuting, while running errands, when going to meet friends and family, driving to a restaurant, etc.
Broadly speaking, it is possible to classify navigation systems as satellite-based~\cite{zaidi2006global}, or inertial based~\cite{weston2000modern}.
Satellite-based navigation systems are aided by, for example, a Global Positioning System (GPS). 
On the other hand, Inertial Navigation Systems (INS) leverage inertial measurement units (IMUs) to maintain the location of a device without the need for any satellite device.
While GPS-based navigation systems generally have greater accuracy than INS, they have shortcomings when it comes to battery life, and sometimes (e.g., indoors where GPS coverage is limited) fail to obtain the GPS lock they require to function.
Unlike satellite-based navigation systems, INS are well suited to applications where either GPS coverage may be limited or where battery life is a concern.

In this work, for the first time, we explore the potential of an \textit{earable}-based inertial navigation system.
Earables (also know as ear-worn wearables) is an exploding area for wearable research. 
They allow users to combine listening to music with sensing and computing.
Specifically to this work, earables offer significant potential benefits for inertial tracking, a key part of an inertial navigation system.
Concretely, earables can be used to track head movements which, in turn, can act as a proxy for visual attention~\cite{ferlini2019head}.
As a result of that, earable could be effectively used as an navigation system by people visually impaired to navigate through audio feedback. 
Additionally, the stabilization effect of the human musculoskeletal postural system ensures a natural vibration damping~\cite{kavanagh2006role} which leads to reduced noise in the motion sensor data fed to the INS.
Furthermore, unlike other wearables, earables present the unique opportunity of two distinct vantage points (one from each ear), which can be leveraged to increase the overall performance of the system.

\begin{figure}[!t]
    \centering
    \includegraphics[width=0.85\linewidth]{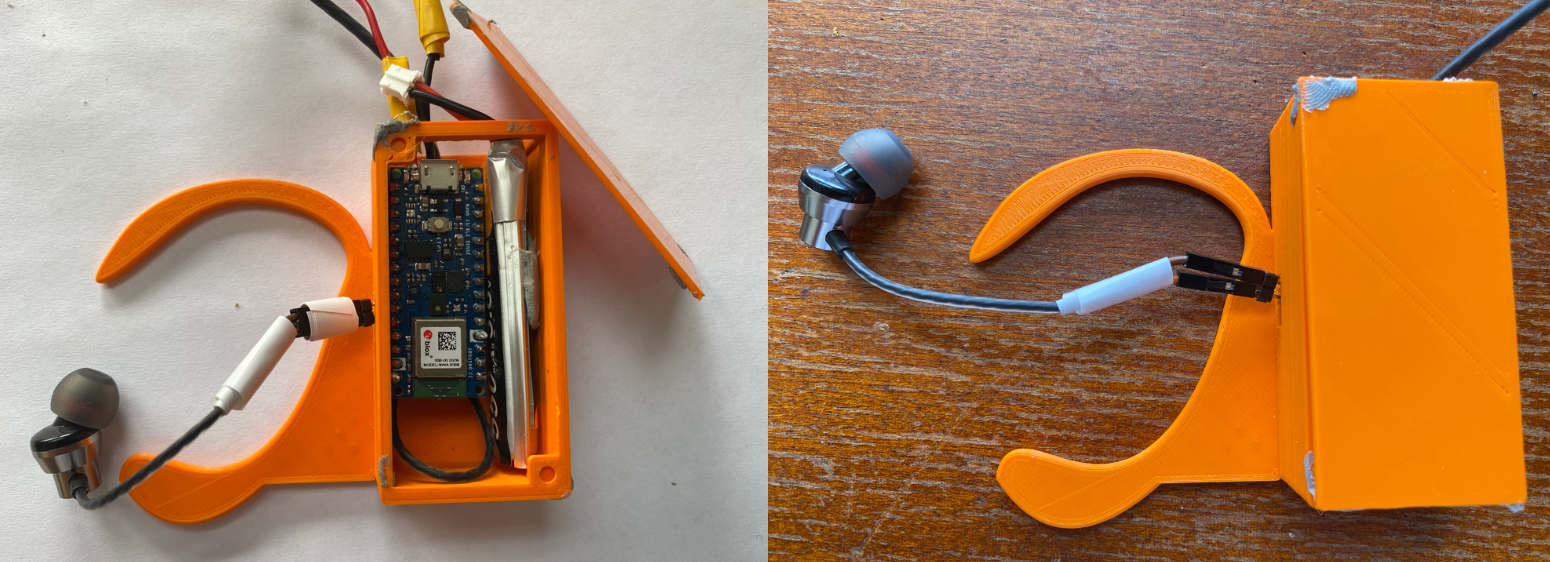}
    \caption{Earable prototype}
    \label{fig:prototype}
\end{figure}

However, to implement an earable-based navigation system, a number of challenges needs to be overcome.
First and foremost the lack of an existing earable platform equipped with a magnetometer~\cite{ferlini2021enabling}: we prototype a new ear-worn sensing platform which leverages a powerful an Arduino Nano 33 BLE Sense as microcontroller (MCU).
Our prototype earable (\Cref{fig:prototype}) collects motion data from the 9-axis IMU on-board the MCU, and streams it over Bluetooth Low Energy (BLE).
Further, it provides audio feedback thanks to a custom-made printed circuit board (PCB) which produces audible tones using pulse width modulation (PWM) and a resistor. 
Our prototype is designed to use minimal power and can run for over a day of continuous use. 
It also contains other sensors (temperature, pressure, microphone) which we do not experiment with in this paper.
Secondly, because of the novelty of earable computing, there is a dataset scarcity.
We therefore run a small user study and collect IMU data using our prototype.
Thirdly, it is well known that inertial approaches require calibration in order to provide accurate results~\cite{grammenos2018you}.
Thus, we build upon previous work to define an effective calibration framework to calibrate the 9 IMU axis of our prototype.
Specifically, we leverage the work done by Sipos et al.~\cite{sipos2011analyses} to calibrate the accelerometer, we implement the gyroscope calibration by Shen et al.~\cite{shen2018closing}, and finally we calibrate the magnetometer following the approach presented by Ferlini et al.~\cite{ferlini2021enabling}.
Lastly, existing position tracking techniques have to be adapted to earables.
We find that a heading estimation approach which leverages the combination of a gyroscopic heading and a magnetometer heading fused together with a complementary filter outperforms existing fusion algorithms.
Correct heading estimation which successfully tracks the position of a person in a navigation application is predicated on displacement estimation.
We perform this estimation by means of a pedestrian dead reckoning algorithm adapted over Lu et al.~\cite{lu2019indoor}.

The contributions of this work can be summarized as follows.
\begin{itemize}
    \item We prototype an earable platform for collecting and transmitting accelerometer, gyroscope, and magnetometer data over Bluetooth Low Energy (BLE) and audio feedback. 
    We use such prototype to collect a new in-the-wild dataset consisting of 9-axis IMU data both from the earable prototype and an iPhone. 
    The user study has been carried out in agreement with the university ethics committee, and comprises data of six users walking both indoors and outdoors, both in low noise and high noise situations. To the best of our knowledge, there is not a similar publicly available dataset, and we will share ours the research community.
    \item We implement and evaluate the performance of PilotEar, the first end-to-end inertial navigation system for earables, achieving an average drift of $0.15\frac{m}{s}$ for one earable and $0.11\frac{m}{s}$ when fusing both earables.
    \item We experiment with a novel multi-device sensor fusion approach to combine both earables and a smartphone. The proposed approach reduces power usage for accurate tracking by 65\% by incorporating occasional GPS updates, whilst reducing tracking drift by 27\% when maximizing performance.
\end{itemize}

\section{Inertial Navigation Primer}\label{sec:background}
Inertial Navigation generally consists of three stages.

The first stage is the \textbf{pre-processing} stage. 
IMU data are sampled and then pre-processed to remove the noise and the temporal drift which often affect IMU readings.
The first challenge is to effectively calibrate each sensor.
In this work, the sensors we consider, and calibrate, are accelerometer, gyroscope, and magnetometer.

Once the IMU data are successfully pre-processed and a calibration framework has been applied, 
Secondly, the pre-processed data is used to find the heading of the individual. This second phase goes by the name \textbf{heading estimation} and, in the case of earables consists in estimating the orientation of the devices.
The heading can be estimated either from the gyroscope~\cite{sola2017quaternion}, from the magnetometer~\cite{ferlini2021enabling}, or fusing the both.
There are many ways to fuse gyroscope and magnetometer data.
In this work we fuse gyroscopic and magnetic data in a similar way to what Shen et al. \cite{shen2018closing} did by mean of a complementary filter. 
Specifically, the gyroscopic and magnetic heading are estimated independently and then are fused as follows:
\begin{equation}
\mathbf{\theta=(1-\alpha) \mathbf{\theta}_{\omega}+\alpha \mathbf{\theta}_{m}}
\end{equation}

\noindent Where ${\theta}_{\omega}$ and ${\theta}_{m}$ are the gyroscopic and magnetic heading, respectively.
The choice of $\alpha$ determines how much of each method is used. Shen et al. \cite{shen2018closing} suggest using a low value to fuse the gyroscope and magnetometer heading. Other papers suggest increasing the value of $\alpha$ over time, as the gyroscopic drift increases.
Although not featured in this paper, other well known fusion algorithms are the Madgwick filter~\cite{madgwick2010efficient}, the Mahony filter~\cite{mahony2005complementary}, and the Fourati filter~\cite{fourati2014heterogeneous}.
We note that our preliminary results suggest the complementary filter approach adapted from Shen et al. \cite{shen2018closing} outperforms our implementation of all the Madgwick, the Mahony and the Fourati filters.

% 3rd stage
Finally, the third and last phase consists in finding the change in position (i.e.,~displacement) of the user at any given timestamp. This phase is also known as \textbf{displacement estimation} phase.
Concretely, the position of the user is estimated by leveraging the their heading and linear acceleration to find the change in their (x, y) position at each timestamp. 
We do that by adapting an existing pedestrian dead reckoning model to complete the step length estimation effectively on earables.
In pedestrian dead reckoning (PDR), the goal is to identify every step that the user makes when walking. 
This is used in practice to identify the distance by which the user has moved at each time step, given the time taken to walk a known length stride.
Our adaptation builds upon the step length estimation proposed by Lu et al.~\cite{lu2019indoor}:
\begin{equation}
    \textbf{D} = K \cdot \iint \boldsymbol{a}(t) d t d t
\end{equation}

\noindent This assumes the velocity for each step starts at $0$, whilst the value of $K$ is computed experimentally for each user.
\section{System Design}\label{sec:sys_design}

The lack of an existing programmable earable platform equipped with a magnetometer compelled us to build to a new earable prototype platform.
Our prototype features a 9-axis IMU, alongside a microphone, a temperature sensor, and pressure sensor. 
The device is easily re-programmable using the Arduino prototyping platform.
We wanted the prototype to be self-contained, thus we included a battery and power management. 
To be useful in real-world applications, the device should also be able to output audio.

\begin{figure}[!h]
    \centering
    \includegraphics[width=0.8\linewidth]{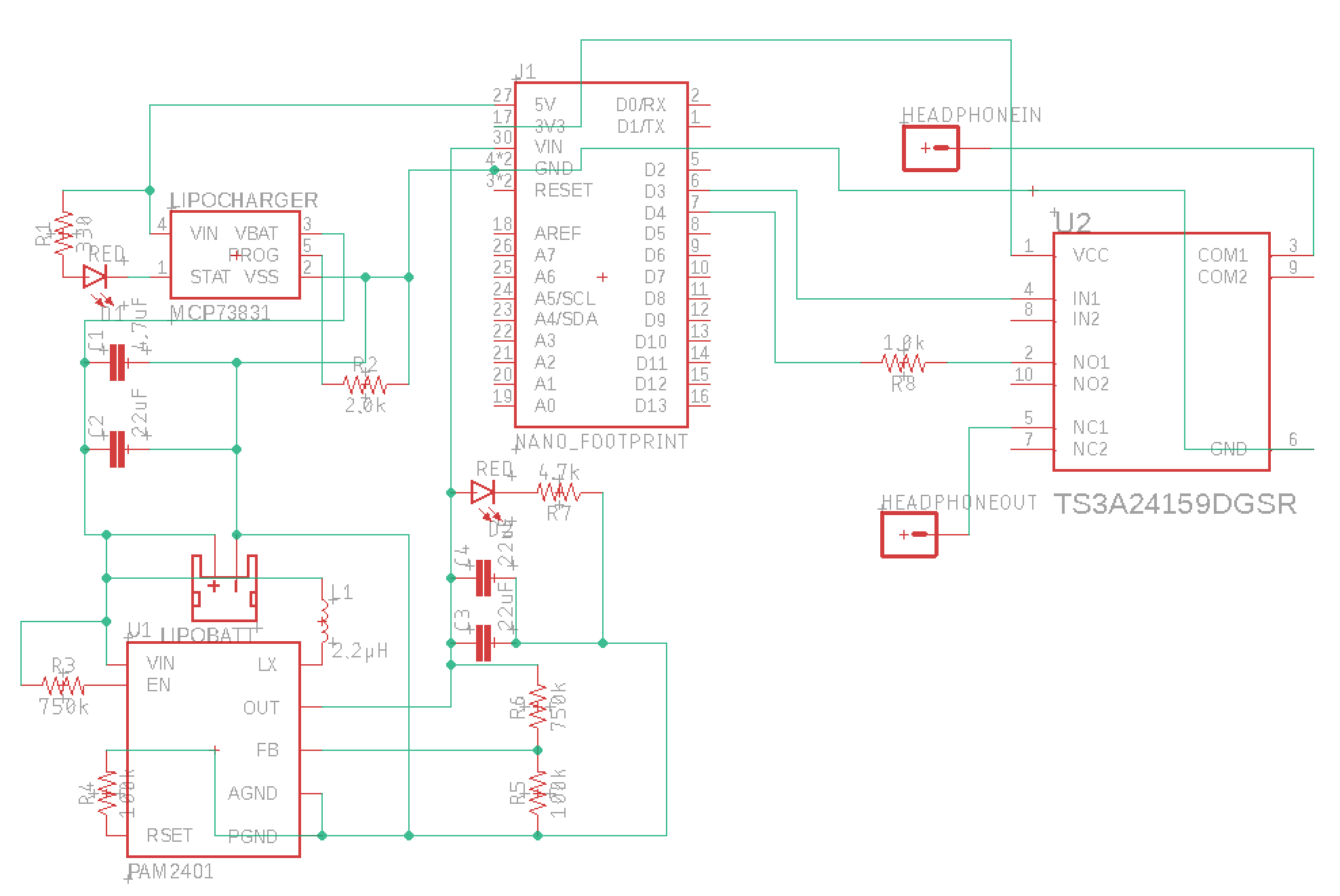}
    \caption{Prototype schematics}
    \label{fig:pcb_design}
\end{figure}

We prototyped our earable platform around an Arduino BLE Sense 33.
This, in turn, runs on a powerful ARM Cortex M4 microcontroller, and is equipped with a BLE transceiver, 9-axis IMU, microphone, temperature and pressure sensors. 
For battery and power management we used a 520mAh lithium polymer battery (LiPo) and power regulation module (PAM2401
%\footnote{\url{https://www.diodes.com/assets/Datasheets/PAM2401.pdf}} 
as a DC-DC boost converter and MCP73831
%\footnote{\url{https://ww1.microchip.com/downloads/en/DeviceDoc/MCP73831-Family-Data-Sheet-DS20001984H.pdf}}
as a charging regulator).
To enable the headphone management and switching circuitry (using a TS3A24159
%\footnote{\url{https://www.ti.com/lit/ds/symlink/ts3a24159.pdf}}
), not natively supported by the Arduino BLE Sense 33, we designed  and printed a two-layer custom PCB (\Cref{fig:pcb_design}).
The PCB design (created using EagleCAD) will be released so that researchers can alter the board for their requirements if necessary.

By using our prototype, researchers can easily collect and transmit data via Bluetooth to a connected device. 
The ubiquitous Arduino platform can be used to write software for this, exploiting the significant existing open-source code. 
We found that we could transmit 9-axis IMU from the prototype earable to a connected an Apple iPhone 11, at 40Hz, with a battery life of over one day (26.3 hours), and a current draw of 20.9mA. 

\begin{figure}[ht]
\centering
\begin{minipage}{.48\linewidth}
  \centering
    \includegraphics[width=.8\linewidth]{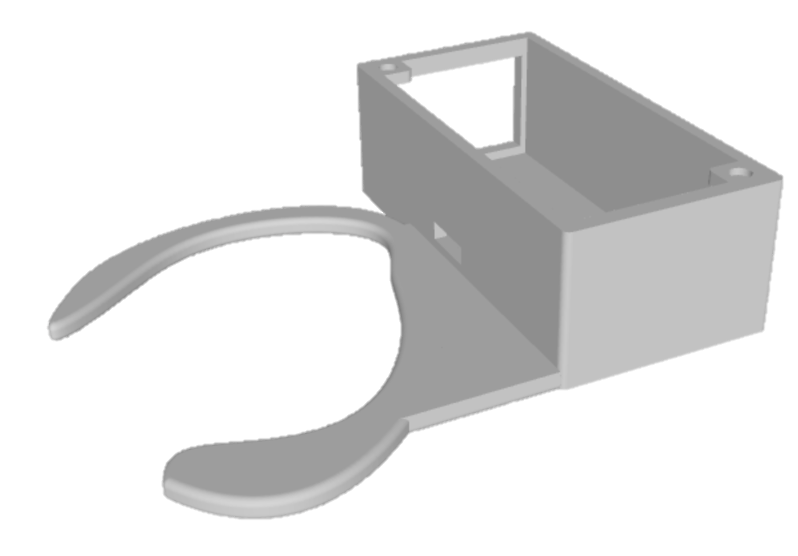}
    \caption{Case design.}
    \label{fig:mk1}
\end{minipage}%
\hfill
\begin{minipage}{.48\linewidth}
  \centering
    \includegraphics[width=.8\linewidth]{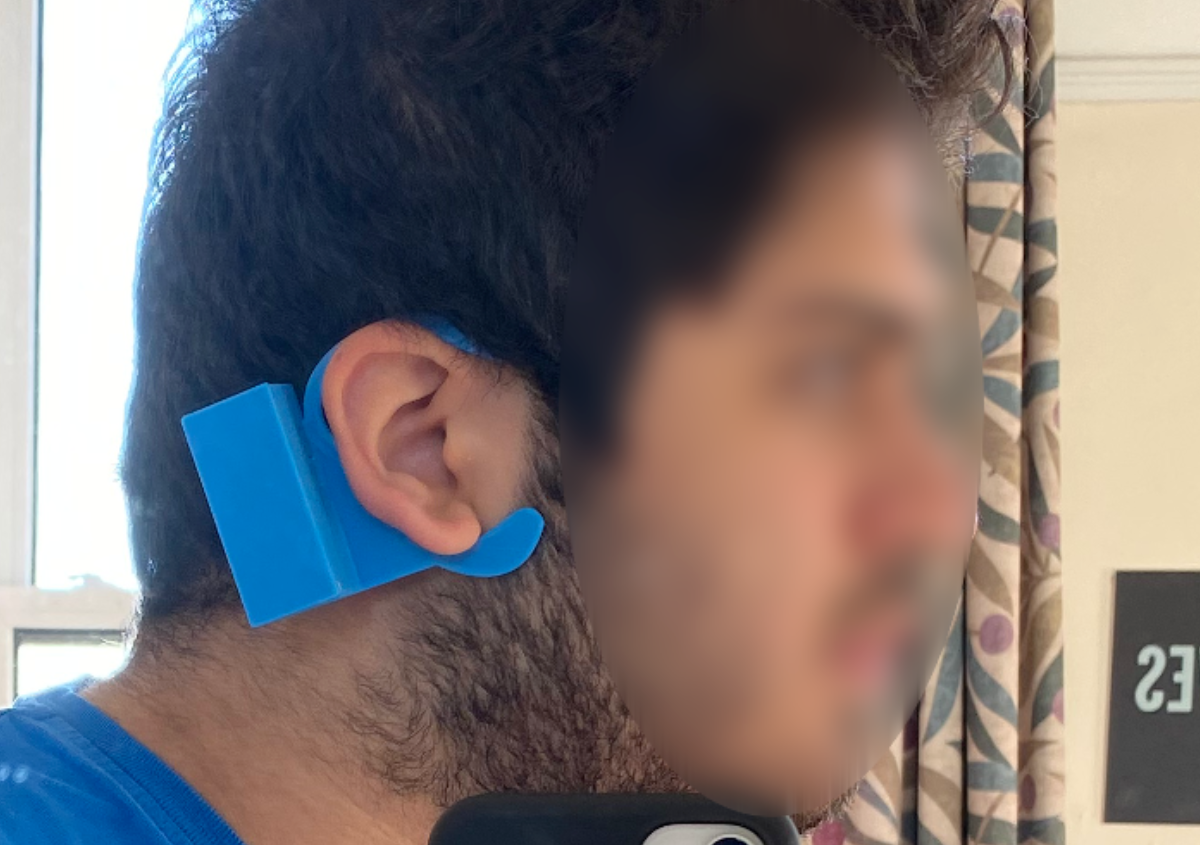}
    \caption{Prototype worn.}
    \label{fig:mi1}
\end{minipage}
\end{figure}

As opposed to other earable platforms, like eSense~\cite{kawsar2018earables}, which are designed as in-ear headphones using injection moulded plastic, our prototype was 3D-printed, with an around-ear design. 
With the current printing material of polylactic acid (PLA), the sizing is user specific (\Cref{fig:mi1}). 
However, if the design could be printed in a more flexible material such as thermoplastic polyurethane (TPU), the design could be more universal. 
\Cref{fig:mk1} shows the prototype design while~\Cref{fig:prototype} shows the complete system which we designed and used to collect data.

\section{Earable Navigation}\label{sec:navigation}
PilotEar consists of a number of elements.
First, sensor calibration.
The aim of this phase is to remove biases and temporal drift from the raw IMU data. 
Following this first step, there is heading estimation, where the calibrate IMU data are used to estimate the direction where the user is facing. 
Since we were only concerned with motion on the plane of the earth's surface, we did not need to deal with any other orientation axis. 
The third aspect aims at estimating the distance moved per time step. 
These three phases, once combined, enable user position tracking from single earable. 
To improve the performance of PilotEar, we also considered several methods to fuse the data sampled by both earables.
Finally, we present a method to provide audio feedback, for instance to a visually impaired user. By doing so, we believe PilotEar could help in guiding a visually impaired person towards the correct exit at a complicated intersection. 

\subsection{Sensor Calibration}
We calibrated the accelerometer using Sipos et al.'s calibration model (\Cref{eq:sipos})~\cite{sipos2011analyses}.
We minimized the square error for calibration clips with each earable in a number of static orientations using the Levenberg-Marquandt Algorithm.

\begin{equation}
\begin{aligned}
\textbf{a}^\prime 
&=\left(\begin{array}{lll}
1 & 0 & 0 \\
\alpha_{y x} & 1 & 0 \\
\alpha_{z x} & \alpha_{z y} & 1
\end{array}\right)\left(\begin{array}{lll}
\mathrm{SF}_{a x} & 0 & 0 \\
0 & \mathrm{SF}_{a y} & 0 \\
0 & 0 & \mathrm{SF}_{a z}
\end{array}\right) 
\times\left(\textbf{a}-\left(\begin{array}{l}
b_{a x} \\
b_{a y} \\
b_{a z}
\end{array}\right)\right)
\end{aligned}
\label{eq:sipos}
\end{equation}

To calibrate the gyroscope, we used a method described by Shen et al.~\cite{shen2018closing}.
Here, when the device is relatively stationary ($|\textbf{acc}| \approx g$), the gyroscope is calibrated using the magnetometer. 
To do so, we integrate the gyroscopic headings over time to find the orientation of the device. 
We then find a rotational offset between the gyroscopic heading and the magnetometer heading.

Finally, for magnetometer calibration, we investigate the real-world tracking efficacy of Ferlini et al.'s semi-automated calibration, which uses occasional phone reference points to calibrate an earable offset~\cite{ferlini2021enabling}. 
This is based on the idea that when a person unlocks and uses their smartphone, the phone and earable are aligned. Hence, the reference heading from the phone (trustworthy) can be used to correct the magnetometer of the earable.
We emulate this by collecting potential calibration points only every $15s$. 
Through preliminary testing, we define a regime to maintain a rolling window of up-to fifteen calibration points, but completing a full re-calibration if we have encountered a full rotation (when the heading rolls over from $359\degree$ to $0\degree$ or vice versa).

\subsection{Heading Estimation}
Our results suggests the best heading estimation comes from simply using the calibrated magnetometer:
\begin{equation}
    \psi = \arctan \frac{m_y}{m_x}
\end{equation}

\noindent To improve the accuracy, we added tilt compensation, to account for the fact that the earth's magnetic field does not run perpendicularly to the surface other than at the equator. 
We do this by finding pitch and roll ($\phi$ and $\theta$, respectively) from the accelerometer, when stationary. 
\begin{equation}
    \phi = \arctan \frac{a_y}{a_z}
\end{equation}
\begin{equation}
    \theta = \arctan \frac{-a_x}{a_y \sin \phi + a_z \cos \phi}
\end{equation}

\noindent $\theta$ and $\phi$ are then used to rotate the magnetometer readings to the flat plane, where $\theta = \phi = 0$

\noindent Whilst offering higher accuracy, the magnetometer-only approach suffers from a potentially lower reliability when a calibration fails.
We, therefore, combined the magnetometer-only heading with the gyroscope-only heading by mean of a complementary filter. $\alpha$ is set to $0.8 - \frac{t}{400}$. 
\begin{multline}
    \psi = (0.8 - \frac{t}{400})\cdot \mathrm{heading\_gyro} + (0.2 + \frac{t}{400})\cdot \mathrm{heading\_mag}
\end{multline}

\noindent Concretely, we trust the gyroscope more at the beginning. 
Over time, the magnetometer calibration is likely to be more refined and, at the same time, the gyroscopic bias may arise. 
Hence the increased trust on the magnetometer.

\subsection{Displacement Estimation}
For the displacement estimation, we adapted Lu et al.'s PDR approach~\cite{lu2019indoor} to identify the stride. We use a constant stride length defined by the height of the user:
\begin{equation}
        \textrm{Stride\_Length} = 0.43 \cdot \textrm{height}
\end{equation}

To find each stride, we used the following algorithm:
\begin{itemize}
    \item Low-pass filter accelerometer norm with 3Hz cut-off frequency.
    \item Find peaks (local maxima) of the filtered data.
    \item Find peaks with topographic prominence above a threshold.
\end{itemize}

\noindent Given the time of each stride, we find the distance moved per timestamp and the change in position.
For each stride occurring between timestamp $i$ and $j$:
\begin{equation}
    d_{t \in i, \dotsc, j} = \frac{\textrm{Step\_Length}}{j-i}
\end{equation}

\noindent Then, for each timestamp:
\begin{equation}
    \textbf{S}_t = \left(\begin{array}{c}
         S_{t_x}  \\
         S_{t_y} 
    \end{array}\right) = \textbf{S}_{t-1} + \left(\begin{array}{c}
         d_t \cos \psi_t  \\
         d_t \sin \psi_t 
    \end{array}\right)
\end{equation}

\subsection{Sensor Fusion}
In combining the data from the two earables, we consider methods to improve the accuracy and reduce the power consumption of the tracking.

To improve the accuracy, we combine the headings of the two earables with a particle filter. 
The complete algorithm for accurate tracking using two earables can be summarized as:
\begin{itemize}
    \item Calibrate both earables independently.
    \item Find the heading at each timestamp for each earable using a complementary filter of gyroscopic and magnetometer heading.
    \item Combine the headings with a particle filter to identify the most probable heading at each timestamp.
    \item Find the stride time according to each earable with Algorithm 1. Average these for both earables.
    \item Find the position according to Equation 10.
\end{itemize}

To reduce the power consumption, we propose two methods. 
First, we consider dropping the sampling frequency of one earable (to 5Hz/ 10Hz), whilst maintaining the frequency of the other at 20Hz. 
We demonstrate that the battery life would increase by 33 minutes when dropping the data frequency from $20Hz$ to $10Hz$, increasing by another 17 minutes going to $5Hz$. 
In the second method, we use GPS updates every 30 seconds.
We take the error model of GPS to be a Gaussian distribution with $\sigma=3.9m$\, since the US claim a 95\% accuracy of $7.8m$ \cite{renfro2018analysis}. 
This is used as the re-sampling probability for a positional particle filter.
This allows us to maintain a maximum average error of $15m$. 
Therefore, by using the current figures from a NEO-6 GPS module, the ability to return the GPS to idle (when a lock is maintained) before using it at high performance for $1s$ would reduce the power usage of a smartphone positional tracking system by 65\%.

\subsection{Audio Feedback}
To demonstrate a potential use-case for tracking using earables, we created an iPhone application to help guide users to a target heading. 
The difference between the calibrated heading and the target heading was found and converted into a tone as follows:
\begin{equation}
    f = 2750(\frac{180-\textrm{heading\_diff}}{180}) + 250
\end{equation}

\noindent By turning their head, the user could audibly gain an understanding of the correct direction, with the tone getting higher in pitch as they are facing the right direction.

\section{Data Collection}\label{sec:dataCollection}
We recruited six healthy subjects (balanced in gender).
The participants wore two earable prototypes, one per ear, and held an iPhone in front of them. 
The system logged 9-axis IMU data and reference heading data from the iPhone.
We collected data while the users walked two different circuits. 
The first was indoors, walking up and down a $10m$ corridor five times. 
The second involved walking outside, where the levels and composition of noise would differ, around a $750m$ route.
In both cases, the test was run twice.
The experiment was granted permission by our university's ethics committee.
\begin{figure}[h!]
  \centering
    \includegraphics[width=.5\linewidth]{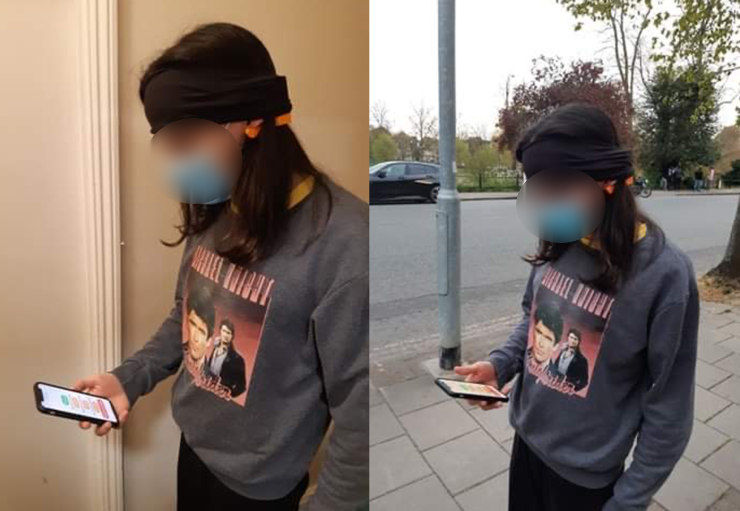}
    \caption{Subject completing the indoor portion of user study (left) and outdoor portion (right).}
    \label{fig:testsubject}
\end{figure}
\section{Evaluation}\label{sec:evaluation}
We test our system on two different metrics. 
The first is the difference between the predicted heading and the ground truth reference heading. 
These are averaged across all the subjects. 
We also look at the drift in tracking position over time (in $\frac{m}{s}$). 
Since GPS could not be used to provide ground truth for the indoor test, both the indoor and outdoor tests were designed to start and end at the same point.
Hence, the drift is the normal of the final predicted position divided by the testing time.

\subsection{Single Earable Tracking}
We first evaluate our calibration framework, showing that the accelerometer calibration, magnetometer calibration and gyroscope calibration methods performed statistically significantly better (using a paired t test with $\alpha = 0.05$) than having no calibration, irrespective of the method used for heading estimation. 
In particular, calibrating the magnetometer leads to a 38\% improvement over no calibration.
We also demonstrate the efficacy of our complementary filter-based heading estimation method, showing how it achieves amongst the lowest heading and displacement errors with lowest standard deviations (\Cref{tab:he_methods}).
This denotes the intuition of weighting the accuracy of magnetometer methods, with better reliability worked well. 
In particular, it statistically significantly outperforms existing sensor fusion algorithms such as Madgwick and Fourati filters.
Further, we compare our displacement estimation method against a simple kinematics approach where the velocity is maintained over time, not exhibiting a velocity drift that was shown by the kinematics method.
\Cref{tab:de_methods} shows how the proposed method outperforms the kinematics one.

\begin{table}[]
\caption{Results comparing heading error for different heading estimation methods.}
\resizebox{.75\linewidth}{!}{%
\begin{tabular}{|ll|l|lll}
\hline
\rowcolor[HTML]{C0C0C0} 
\textbf{} & \textbf{} & \multicolumn{2}{l|}{\cellcolor[HTML]{C0C0C0}\textbf{Heading Error / \degree}} & \multicolumn{2}{l}{\cellcolor[HTML]{C0C0C0}\textbf{Displacement Error / $ms^{-1}$}} \\ \cline{3-6} 
\rowcolor[HTML]{C0C0C0} 
\textbf{} & \textbf{} & \textbf{Average} & \multicolumn{1}{l|}{\cellcolor[HTML]{C0C0C0}\textbf{\begin{tabular}[c]{@{}l@{}}Standard\\ Deviation\end{tabular}}} & \multicolumn{1}{l|}{\cellcolor[HTML]{C0C0C0}\textbf{Average}} & \multicolumn{1}{l|}{\cellcolor[HTML]{C0C0C0}\textbf{\begin{tabular}[c]{@{}l@{}}Standard\\ Deviation\end{tabular}}} \\ \hline
\multicolumn{2}{|l|}{\cellcolor[HTML]{C0C0C0}\textbf{Magnetometer only}} & 15.1 & \multicolumn{1}{l|}{15.7} & \multicolumn{1}{l|}{0.133} & \multicolumn{1}{l|}{0.0354} \\ \hline
\multicolumn{2}{|l|}{\cellcolor[HTML]{C0C0C0}\textbf{Gyroscope only}} & 21.9 & \multicolumn{1}{l|}{15.7} & \multicolumn{1}{l|}{0.194} & \multicolumn{1}{l|}{0.0217} \\ \hline
\multicolumn{2}{|l|}{\cellcolor[HTML]{C0C0C0}\textbf{Madgwick Filter}} & 19.5 & \multicolumn{1}{l|}{15.7} & \multicolumn{1}{l|}{0.228} & \multicolumn{1}{l|}{0.0423} \\ \hline
\multicolumn{2}{|l|}{\cellcolor[HTML]{C0C0C0}\textbf{Fourati Filter}} & 21.9 & \multicolumn{1}{l|}{16} & \multicolumn{1}{l|}{0.181} & \multicolumn{1}{l|}{0.0275} \\ \hline
\multicolumn{2}{|l|}{\cellcolor[HTML]{C0C0C0}\textbf{\begin{tabular}[c]{@{}l@{}}Complementary\\ Filter\end{tabular}}} & 15.8 & \multicolumn{1}{l|}{14.0} & \multicolumn{1}{l|}{0.132} & \multicolumn{1}{l|}{0.0214} \\ \hline
\end{tabular}%
}

\label{tab:he_methods}
\end{table}

\begin{table}[]
\caption{Results comparing displacement error for different displacement estimation methods.}
\resizebox{.96\linewidth}{!}{%
\begin{tabular}{|l|l|l|l|l|l|l|l|}
\hline
\rowcolor[HTML]{C0C0C0} 
\multicolumn{2}{|l|}{\cellcolor[HTML]{C0C0C0}} & \multicolumn{2}{l|}{\cellcolor[HTML]{C0C0C0}} & \multicolumn{2}{l|}{\cellcolor[HTML]{C0C0C0}\textbf{Kinematics}} & \multicolumn{2}{l|}{\cellcolor[HTML]{C0C0C0}\textbf{Step Length Estimation}} \\ \cline{5-8} 
\rowcolor[HTML]{C0C0C0} 
\multicolumn{2}{|l|}{\cellcolor[HTML]{C0C0C0}} & \multicolumn{2}{l|}{\multirow{-2}{*}{\cellcolor[HTML]{C0C0C0}\textbf{Heading Error / \degree}}} & \multicolumn{2}{l|}{\cellcolor[HTML]{C0C0C0}\textbf{\begin{tabular}[c]{@{}l@{}}Displacement Error\\ / $ms^{-1}$\end{tabular}}} & \multicolumn{2}{l|}{\cellcolor[HTML]{C0C0C0}\textbf{\begin{tabular}[c]{@{}l@{}}Displacement Error\\ / $ms^{-1}$\end{tabular}}} \\ \cline{3-8} 
\rowcolor[HTML]{C0C0C0} 
\multicolumn{2}{|l|}{\multirow{-3}{*}{\cellcolor[HTML]{C0C0C0}}} & \textbf{Average} & \textbf{\begin{tabular}[c]{@{}l@{}}Standard\\ Deviation\end{tabular}} & \textbf{Average} & \textbf{\begin{tabular}[c]{@{}l@{}}Standard\\ Deviation\end{tabular}} & \textbf{Average} & \textbf{\begin{tabular}[c]{@{}l@{}}Standard\\ Deviation\end{tabular}} \\ \hline
\multicolumn{2}{|l|}{\cellcolor[HTML]{C0C0C0}\textbf{Magnetometer only}} & 15.1 & 15.6 & 2.32 & 0.769 & 0.133 & 0.0354 \\ \hline
\multicolumn{2}{|l|}{\cellcolor[HTML]{C0C0C0}\textbf{Gyroscope only}} & 21.9 & 15.7 & 2.69 & 1.11 & 0.194 & 0.0217 \\ \hline
\multicolumn{2}{|l|}{\cellcolor[HTML]{C0C0C0}\textbf{Madgwick Filter}} & 19.4 & 15.7 & 3.66 & 1.55 & 0.228 & 0.0423 \\ \hline
\multicolumn{2}{|l|}{\cellcolor[HTML]{C0C0C0}\textbf{Fourati Filter}} & 21.9 & 16 & 2.42 & 1.14 & 0.181 & 0.0275 \\ \hline
\multicolumn{2}{|l|}{\cellcolor[HTML]{C0C0C0}\textbf{Complementary Filter}} & 15.8 & 14.0 & 2.63 & 0.538 & 0.132 & 0.0214 \\ \hline
\end{tabular}%
}

\label{tab:de_methods}
\end{table}

\begin{figure}[!h]
        \centering
        \includegraphics[width=0.75\linewidth]{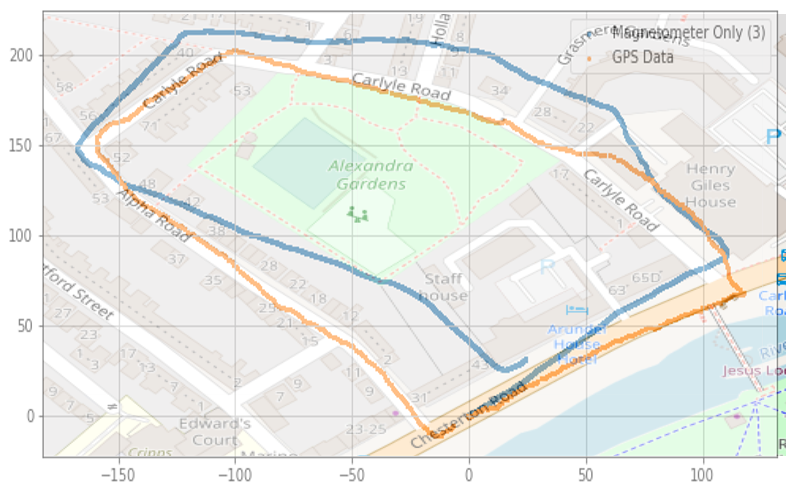}
        \caption{Example of outdoor tracking path.}
        \label{fig:pe_methods2}
\end{figure}

\Cref{fig:pe_methods2} reports an example path found by tracking a single earable on the outdoor test, showing how our method for single earable tracking performed very well, with a minimum average tracking drift of around $0.15\frac{m}{s}$.

\subsection{Two Earable Tracking}

\begin{table}[]
\caption{Results for combining both earables when particle filtering the heading.}
\resizebox{.68\linewidth}{!}{%
\begin{tabular}{|
>{\columncolor[HTML]{C0C0C0}}l |l|l|l|l|}
\hline
\cellcolor[HTML]{C0C0C0} & \multicolumn{2}{l|}{\cellcolor[HTML]{C0C0C0}\textbf{Indoors}} & \multicolumn{2}{l|}{\cellcolor[HTML]{C0C0C0}\textbf{Outdoors}} \\ \cline{2-5} 
\cellcolor[HTML]{C0C0C0} & \multicolumn{2}{l|}{\cellcolor[HTML]{C0C0C0}\textbf{\begin{tabular}[c]{@{}l@{}}Final Displacement \\ Error / $ms^{-1}$\end{tabular}}} & \multicolumn{2}{l|}{\cellcolor[HTML]{C0C0C0}\textbf{\begin{tabular}[c]{@{}l@{}}Final Displacement\\ Error / $ms^{-1}$\end{tabular}}} \\ \cline{2-5} 
\multirow{-3}{*}{\cellcolor[HTML]{C0C0C0}} & \cellcolor[HTML]{C0C0C0}\textbf{Average} & \cellcolor[HTML]{C0C0C0}\textbf{\begin{tabular}[c]{@{}l@{}}Standard\\ Deviation\end{tabular}} & \cellcolor[HTML]{C0C0C0}\textbf{Average} & \cellcolor[HTML]{C0C0C0}\textbf{\begin{tabular}[c]{@{}l@{}}Standard\\ Deviation\end{tabular}} \\ \hline
\textbf{Left Earable} & 0.173 & 0.096 & 0.181 & 0.025 \\ \hline
\textbf{Right Earable} & 0.135 & 0.081 & 0.170 & 0.082 \\ \hline
\textbf{Reference Heading} & 0.077 & 0.038 & 0.104 & 0.061 \\ \hline
\textbf{Both Earables} & 0.119 & 0.089 & 0.021 & 0.039 \\ \hline
\end{tabular}%
}

\label{tab:experiment1}
\end{table}

\begin{figure}[!h]
    \centering
    \includegraphics[width=.7\linewidth]{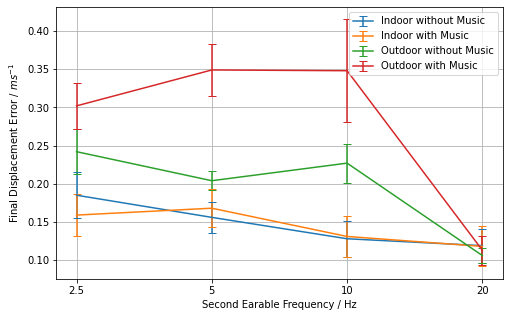}
    \caption{Tracking drift when mixing timings of earables.}
    \label{fig:exp4_graph}
\end{figure}

\Cref{tab:experiment1} demonstrates how combining the data from both earables statistically significantly outperforms using one earable. 
This difference is more notable outside, likely due to the additional noise.
Our results (\Cref{tab:experiment_4} and \Cref{fig:exp4_graph}) suggest that dropping the data frequency of one earable generally adversely impacted the displacement error. 
For indoor tests, there was a statistically significant difference going from $20Hz$ to $10Hz$ and $10Hz$ to $5Hz$, however, there was no difference going from $5Hz$ to $2.5Hz$. 
Additionally, the results for $10Hz$ and $5Hz$ still performed better  than one earable individually. 
For the outdoor tests, the changes in data frequency had a less obvious effect. 
There was no clear trend with the reducing the second earable's frequency. However, for all frequency reductions, there was a significant negative impact. 
We suspect that this arose from an increased background noise that came from the outdoor environment. 
This noise meant that additional measurements were important for tracking accuracy.

\begin{table}[]
\caption{Results for mixing timings for earables.}
\resizebox{.68\linewidth}{!}{%
\begin{tabular}{|
>{\columncolor[HTML]{C0C0C0}}l |l|l|l|l|}
\hline
\cellcolor[HTML]{C0C0C0} & \multicolumn{2}{l|}{\cellcolor[HTML]{C0C0C0}\textbf{Indoors}} & \multicolumn{2}{l|}{\cellcolor[HTML]{C0C0C0}\textbf{Outdoors}} \\ \cline{2-5} 
\cellcolor[HTML]{C0C0C0} & \multicolumn{2}{l|}{\cellcolor[HTML]{C0C0C0}\textbf{\begin{tabular}[c]{@{}l@{}}Final Displacement\\ Error / $ms^{-1}$\end{tabular}}} & \multicolumn{2}{l|}{\cellcolor[HTML]{C0C0C0}\textbf{\begin{tabular}[c]{@{}l@{}}Final Displacement\\ Error / $ms^{-1}$\end{tabular}}} \\ \cline{2-5} 
\multirow{-3}{*}{\cellcolor[HTML]{C0C0C0}\textbf{}} & \cellcolor[HTML]{C0C0C0}\textbf{Average} & \cellcolor[HTML]{C0C0C0}\textbf{\begin{tabular}[c]{@{}l@{}}Standard\\ Deviation\end{tabular}} & \cellcolor[HTML]{C0C0C0}\textbf{Average} & \cellcolor[HTML]{C0C0C0}\textbf{\begin{tabular}[c]{@{}l@{}}Standard\\ Deviation\end{tabular}} \\ \hline
\textbf{Left Earable} & 0.173 & 0.096 & 0.226 & 0.025 \\ \hline
\textbf{Right Earable} & 0.135 & 0.081 & 0.139 & 0.082 \\ \hline
\textbf{Reference Heading} & 0.077 & 0.038 & 0.168 & 0.061 \\ \hline
\textbf{Both 20Hz} & 0.119 & 0.089 & 0.106 & 0.039 \\ \hline
\textbf{One 20Hz, one 10Hz} & 0.128 & 0.094 & 0.227 & 0.102 \\ \hline
\textbf{One 20Hz, one 5Hz} & 0.156 & 0.081 & 0.204 & 0.049 \\ \hline
\textbf{One 20Hz, one 2.5Hz} & 0.185 & 0.119 & 0.242 & 0.119 \\ \hline
\end{tabular}%
}

\label{tab:experiment_4}
\end{table}
\section{Related Work}\label{sec:related_work}
Earable research is living a momentum~\cite{choudhury2021earable}.
In the literature earables have been used for a number of applications ranging form head motion tracking~\cite{ferlini2019head, ferlini2021enabling}, to inertial sensing for activity recognition~\cite{kawsar2018earables} and step counting~\cite{prakash2019stear}, acoustic sensing~\cite{ma2021oesense}, and a number of mobile health applications.
A few (non comprehensive list of) examples are dietary habits monitoring~\cite{amft2005analysis, bedri2017earbit, bi2018auracle}, teeth grinding (Bruxism)/jaw clenching~\cite{rupavatharam2019towards} detection, body-core temperature and blood pressure monitoring~\cite{bui2019ebp}, in-ear photoplethysmography (PPG)~\cite{passler2019ear, patterson2009flexible}, and sleep monitoring~\cite{nguyen2016lightweight}.
In this work, for the first time, we leverage earables to build an end-to-end navigation system.

There is a significant body of related work for inertial navigation. 
This includes papers by Shen et al. \cite{shen2018closing}, Woodman et al. \cite{woodman2007introduction}, and Kok et al. \cite{kok2017using} who create full INS using 9-axis IMUs. 
There are also significant works which focus on specific components of the process, some for wearable devices. 
Won et al. \cite{won2009triaxial}, Frosio et al. \cite{frosio2012autocalibration} and Skog et al. \cite{skog2006calibration} look at accelerometer calibration. 
For gyroscope calibration, there is significant work on calibration methods that require precise turntables, including work by Yang et al. \cite{yang2018novel} and Chen et al. \cite{chen2018chip}. 
Finally, for magnetometer calibration, there is work by Renaudin et al. \cite{renaudin2010complete}. 
We instead use the method defined for earables by Ferlini et al. \cite{ferlini2021enabling}. 
For heading estimation, there are numerous approaches, some of which are considered in this paper. 
These include methods by Madgwick et al. \cite{madgwick2010efficient} and Fourati et al. \cite{fourati2014heterogeneous}. 
Other prominently used methods include Mahony et al. \cite{mahony2005complementary} and QUEST \cite{crassidis2007survey}. 
For displacement estimation, there is previous work on methods which work well for wearables, using features of hand movement~\cite{shen2018closing}, and foot movement~\cite{jimenez2009comparison} to improve performance. 
However, to the best of our knowledge, there is no existing work which carries out any inertial navigation on earables.
\section{Final Remarks}\label{sec:conclusion}
In this work we introduce PilotEar, a novel framework for earable-based inertial navigation. 
PilotEar is capable of achieving a drift of as little as $0.11 \frac{m}{s}$, when fusing the data coming from both ears.
Our work defines an effective calibration schema, as well as precise heading and displacement estimation methods. 
Ultimately, we believe that PilotEar, providing acoustic feedback, could offer significant benefits for navigation, particularly for visually impaired people.

\begin{acks}
This work is supported by Nokia Bell Labs through their donation for the Centre of Mobile, Wearable Systems and Augmented Intelligence.
\end{acks}


\end{document}